\documentclass[english,pre,twocolumn,floatfix]{revtex4}
\usepackage{graphicx}  
\usepackage{dcolumn}   
\usepackage{bm}        
\usepackage{amssymb}   
\usepackage{amsmath}

\usepackage[T1]{fontenc}
\usepackage[latin9]{inputenc}
\usepackage{amsthm}
\usepackage{amsmath}
\usepackage{graphicx}

\makeatletter
\@ifundefined{textcolor}{}
{%
 \definecolor{BLACK}{gray}{0}
 \definecolor{WHITE}{gray}{1}
 \definecolor{RED}{rgb}{1,0,0}
 \definecolor{GREEN}{rgb}{0,1,0}
 \definecolor{BLUE}{rgb}{0,0,1}
 \definecolor{CYAN}{cmyk}{1,0,0,0}
 \definecolor{MAGENTA}{cmyk}{0,1,0,0}
 \definecolor{YELLOW}{cmyk}{0,0,1,0}
 }

\makeatletter

\newcommand{\Rmnum}[1]{\expandafter\@slowromancap\romannumeral #1@}
\makeatother

\newcommand{\be}{\begin{equation}}
\newcommand{\ee}{\end{equation}}

\newcommand{\kucb}{k^{\rm cb}_u}
\newcommand{\kusb}{k^{\rm sb}_u}

\newcommand{\kb}{k_{\rm B}}

\newcommand{\Nt}{N_{\rm t}}
\newcommand{\Ft}{F_{\rm t}}

\newcommand{\kBT}{{k_{\textrm{B}} T}}

\newcommand{\dx}[1]{\mathrm{d} #1}
\newcommand{\DDx}[2]{\frac{\dx{#1}}{\dx{#2}}}

\def\lsim{\mathrel{\rlap{\lower4pt\hbox{$\sim$}}
    \raise1pt\hbox{$<$}}}                
\def\gsim{\mathrel{\rlap{\lower4pt\hbox{$\sim$}}
    \raise1pt\hbox{$>$}}}

\renewcommand\[{\begin{equation}}
\renewcommand\]{\end{equation}}
\usepackage[english]{babel}
\addto\captionsenglish{}
\makeatother

\usepackage{babel}

\begin{document}

\title{Evolving roles and dynamics for catch and slip bonds during adhesion cluster maturation}

\author{Elizaveta A. Novikova$^{1,2}$ and Cornelis Storm$^{2,3}$}

\affiliation{$^{1}$Institut de Biologie de l'Ecole Normale Superieure (IBENS), D\'epartement de Biologie, Ecole Normale Sup\'erieure, CNRS, Inserm, PSL Research University, 46 rue d'Ulm, 75005 Paris, France\\$^{2}$Department of Applied Physics, Eindhoven University of Technology, P. O. Box 513,
NL-5600 MB Eindhoven, The Netherlands\\$^{3}$Institute for Complex
Molecular Systems, Eindhoven University of Technology, P. O. Box 513,
NL-5600 MB Eindhoven, The Netherlands}

\date{\today}
\begin{abstract}
Focal adhesions are the loci of cellular adhesion to the extracellular matrix. At these sites, various integrins forge connections between the intracellular cytoskeleton and the outside world: large patches of multiple types of integrins together grip hold of collagen, fibronectin and other extracellular matrix components. The mixture of integrins composing the FA will, in general, contain both slip bond integrins and catch bond integrins---bonds whose lifetime increases with applied load and bonds for whom it decreases when forced. Prior work suggests that catch bonds are essential for proper FA stability and mechanosensory functionality. In the present work, we investigate, numerically, the interplay between the two distinct types of bonds and ask how the presence, in the same FA cluster, of slip bonds augments the behavior of the catch bonds. We show, that mixing the two components m low-force mechanical integrity, lacking in purely catch systems, while preserving the potential to strengthen the FA bond by force as well as the mechanosensory qualities of the catch bonds. We investigate the spatial distribution in mixed-integrin FA's and show that the differential response to loading leads, via an excluded volume interaction, to a dependence of the individual integrin diffusivities on the applied load, an effect that has been reported in experiments.
\end{abstract}
\maketitle

\section{Introduction}
 Cells are able to sense and react to the stiffness of the extracellular environment \cite{lo2000cell,Discher18112005,kong2005non}. Through their focal adhesions (FAs) cells are able to confer mechanical forces onto the extracellular matrix (ECM). Inside each FA \cite{kanchanawong2010nanoscale} transmembrane proteins called integrins provide direct links between the cells' internal contractile machinery and various ECM components.  Integrins are heterodimers, composed of two subunits called $\alpha$ and $\beta$; each of these comes in various kinds. Together, there are about 25 different integrins in the vertebrates, allowing their cells to form robust adhesions to ECM components like collagen, fibronectin and vitronectin. Within a single FA, multiple integrin species are generally represented \cite{cluzel2005mechanisms} and previous work suggests that this mixed nature of FA provides enhanced functionality. For instance, the signaling pathways of two widely studied and commonly cohabitating integrin types--- $\alpha_5 \beta_1$ and $\alpha_V \beta_3$---interfere \cite{white2007alphavbeta3}, and their roles in adhesion and motility complement each other \cite{Roca-Cusachs22092009, Balcioglu1316}. Interactions---direct or indirect---between integrins of different types have been implicated in guiding force generation and rigidity sensing \cite{RocaCusachs2014}. To better understand the roles of integrins in adhesion, models of the mechanosensing and mechanotransduction mechanisms on the level of the cell \cite{Walcott27042010, MarcqActiveMatter} and on the level of focal adhesion \cite{Schwarz2006225, PhysRevLett.92.108102,Schwarz2007,RocaCusachs2014} were developed. In this work we consider a focal adhesion with two types of integrins, a system analogous to the one that was  treated in \cite{RocaCusachs2014}.  We complete the findings  in \cite{RocaCusachs2014} with the analysis of mixed cluster stability and a modeling of integrin mobility inside the focal adhesion.  Our results exploit and extend our simulations in \cite{Novikova2013}: We supply a more realistic model, we develop a method to determine macroscopic stiffness-dependent parameters of the focal adhesion. The central question that we answer here is: How does the force, exerted on focal adhesion, influence the diffusivity of free integrins inside it? We consider a force-response of a focal adhesion consisting of two integrin types under a constant force load. Using the assumption of uniform load sharing, we determine the equilibrium values of such a mixed cluster, depending on the individual properties of the bonds in it, and on the cluster composition. We explore the stability of a mixed cluster under load and then include simple lateral diffusion of integrins on a two-dimensional lattice. We determine the diffusivity of free (unbound) integrins cas a function of the applied force, and suggest that the diffusivity of integrins inside a focal adhesion is a macroscopic parameter which reflects the force exerted on it. Assuming that a cell invests equal energy in every focal adhesion, we conclude that the diffusivity of integrins inside a focal adhesion depends on the stiffness of the extracellular matrix and the level of force, which increases as the adhesion matures.
\section{Binding and unbinding of single catch and slip bonds}
The binding and unbinding rates $k_b$ and $k_u$ characterize the equilibrium kinetics of a single, noncovalent  molecular bond. These rates are load-dependent; in response to an applied pulling force $f$ the unbinding rate of so-called \emph{slip-bonds} is predicted, according to Kramer's rate theory \cite{Kramers}, to increase exponentially as
\be\label{slipbare}
\kusb=k_0^{\rm sb}\exp{\left(\frac{+f \xi_{\rm sb}}{\kBT}\right)}\, .
\ee
In this expression, $\xi_{\rm sb}$ is a microcscopic unbinding length; $\kb$ is the Boltzmann constant and $T$ is the absolute temperature. $k_0^{\rm sb}$ is the unforced unbinding rate; the rate at which the bond opens up under the effect of spontaneous fluctuations. It is set by a bare attempt frequency $k_0$ and $\Delta U_{\rm sb}$, the height of the energetic barrier corresponding to the dissociation of the bond
\be
k_0^{\rm sb}=k_0 \exp{\left(\frac{\Delta U_{\rm sb}}{\kBT}\right)}\, .
\ee
In the case of a \emph{catch-bond}, the unbinding behavior \cite{Dembo} is quite different: When a moderate tension is applied to this bond, the bond dissociation rate initially {\em decreases}, corresponding to an increase in the single-bond lifetime. Using the so-called 'two pathway model' \cite{PereverzevTwoPathway}, the total unbinding rate of such a catch bond may be computed as 
\be\label{catchbare}
\kucb=k_{0,1}^{\rm cb}\exp{\left(\frac{+f \xi_1}{\kBT}\right)}+k_{0,2}^{\rm cb}\exp{\left(\frac{-f \xi_2}{\kBT}\right)}\, ,
\ee
that is, a sum of two rates of corresponding to two parallel processes. Process 1 (with bare unbinding rate $k_{0,1}$ and dissociation length $\xi_1$) captures dissociation along a slip-like path, as may be surmised from the increase in rate with increasing force. Process 2 (with bare unbinding rate $k_{0,2}$ and dissociation length $\xi_2$) describes dissociation along a catch path, different in the sense that the force-dependence in the exponent carries a minus sign which leads to a decreasing catch unbinding rate.

\begin{figure}[th]
\includegraphics[width=\columnwidth]{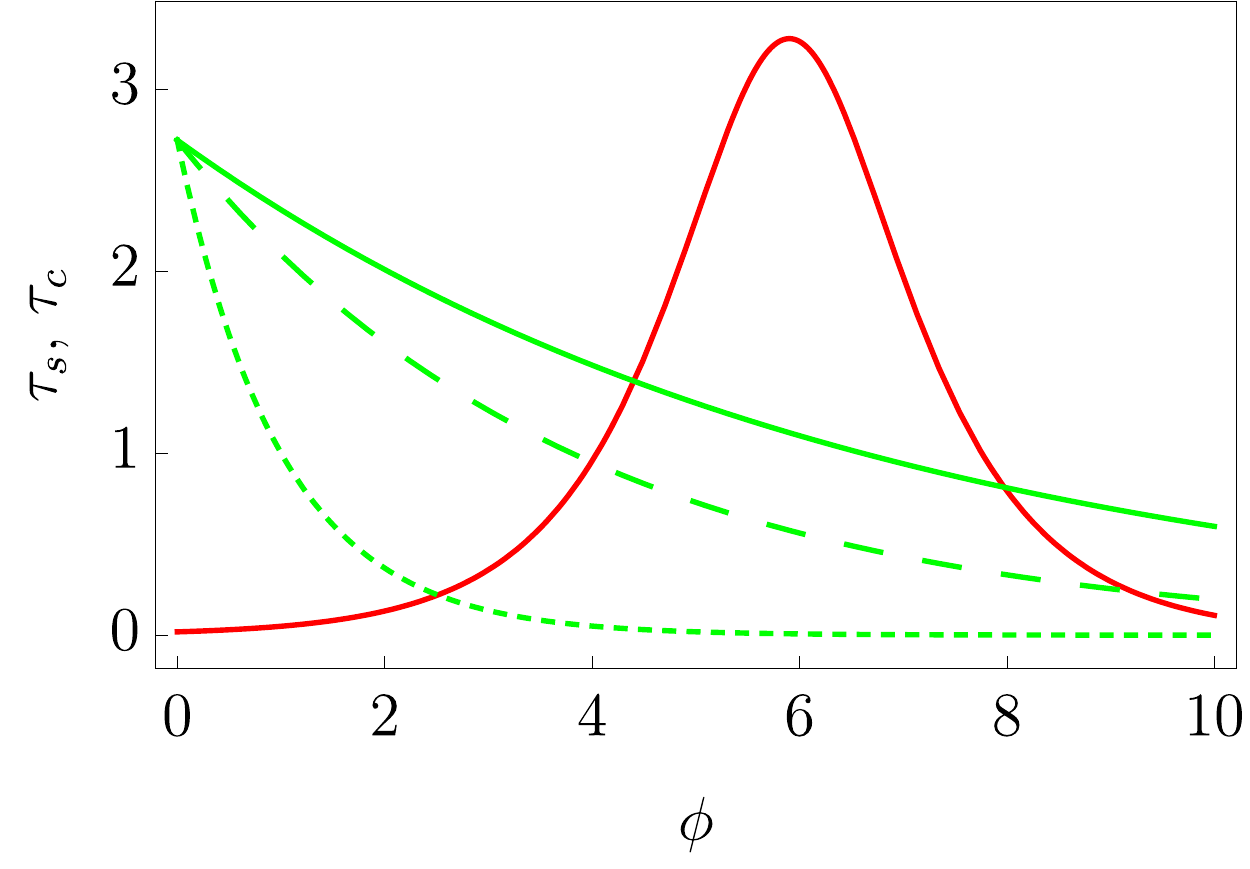}\caption{\label{Fig1}Average lifetimes $\tau_c,\tau_s$ and  of catch- (red) and slip (green)  bonds as a function of dimensionless force, $\phi$. Parameter values for catch bonds (described by  Eq.\,(\ref{cbf})): $\phi_2=4.02$, $\phi_1=7.78$; for slip bonds (described by Eq.\,(\ref{sbf})) short dashed: $\rho_{\xi}=1$, $u_{\rm sb}=1$, long dashed: $\rho_{\xi}=3.8$,$u_{\rm sb}=1$, solid line: $\rho_{\xi}=6.6$, $u_{\rm sb}=1$. }
\end{figure}

In previous work \cite{Novikova2013}, we show how to re-express Eq.\,(\ref{catchbare}) in terms of the normalized catch bond unbinding rate (in the case that $\xi_1=\xi_2\equiv \xi_{\rm cb}$ as a function of the dimensionless force $\phi={f \xi_{\rm cb}}/{\kBT}$ using only two parameters $\phi_1$ and $\phi_2$ reflecting, respectively, the dissociation energy barriers for the slip- and the catch path.:
\be\label{cbf}
\kucb(\phi)=e^{(\phi-\phi_1)}+e^{-(\phi-\phi_2)}\, .
\ee 
In the present work, we are interested in coupling these catch bonds with slip bonds. Their unbinding rate Eq.\,(\ref{slipbare}), likewise, may be re-expressed in terms of the same nondimensional forces and rates
\be\label{sbf}
\kusb(\phi)=e^{(\phi / \rho_{\xi}-u_{\rm sb})}\, ,
\ee
with two additional parameters: $\rho_\xi=\xi_{\rm cb}/\xi_{\rm sb}$ is the ratio of the catch and slip bond dissociation lengths, and $u_{\rm sb}=-\Delta U_{\rm sb}/{\kBT}$ sets the zero-force unbinding rate of the slip bond. Once its unbinding rate $k_u$ is known, the average lifetime of a single bond is computed as
\be
k_u(\phi)\equiv(k_0 \tau(\phi))^{-1}\, .
\ee
After their discovery, single biological catch-bonds have received considerable attention in the community. Recent experiments \cite{DirectObservationOfC, Kong29062009, C2SM07171A} measured catch-bond characteristics by pulling single integrin-ligand bond with an AFM-tip. In this work we use the parameters of an individual integrin- fibronectin catch bond, which were obtained in one of these experiments \cite{DirectObservationOfC}. As earlier in \cite{Novikova2013}, we use the two-pathway model from \cite{PereverzevTwoPathway}, and fit it to the data from \cite{Kong29062009}, with fit parameters $\phi_1$ and $\phi_2$. The dimensionless force $\phi$ is computed as $\phi=f/f^\star$, where $f^\star=5.38$ is a scaling force. As also noted in \cite{RocaCusachs2014}, compared to catch-bonds slip-bonds formed by integrins have not been studied in as much detail; for demonstrational purposes we will, in the present paper, fix the catch bond parameters at the aforementioned values and will vary the slip bond parameter $\rho_\xi$ to set the relative force responsivity. Throughout this paper, we set $u_{\rm sb}=1$ as the reference, zero-force unbinding rate for slip bonds. In Fig. \ref{Fig1}, we plot the resulting catch- and slip lifetimes for various values of $\rho_\xi$. The distinct force-lifetime responses are clearly visible with the catch bond showing the characteristic maximum at finite force at the unbinding lifetime. 
 With these preliminaries in place, we turn to the behavior of a mixed cluster containing both catch and slip bonds, at finite force.
\section{A mixed catch-slip cluster at fixed force: Mean field theory} 

Following the approach laid out in Schwarz {\em et al.} \cite{Schwarz2006225}, we consider a fixed total number of bonds (bound or unbound) $\Nt$, out of which $N_{\rm ct}$ are catch-bonds and $N_{\rm st}$ are slip-bonds; $N_{\rm ct}$ and $N_{\rm st}$ are individually conserved. We will let $i$ denote the number of bound catch bonds, and $j$ the number of bound slip bonds at time $t$. We denote the probability of having $i$ closed catch bonds and $j$ closed slip bonds at a given time $t$ by $p_{i,j}(t)$; its evolution is governed by a a one-step, two-variate master equation:
\begin{eqnarray}\label{lsme}
\DDx{p_{i,j}(t)}{t}&=&r^s_{i,j+1}(\Ft)p_{i,j+1}+r^c_{i+1,j}(\Ft)p_{i+1,j}\nonumber\\
&&+g^c_{i-1,j}p_{i-1,j}+g^s_{i,j-1}p_{i,j-1}\nonumber\\ &&-\left[r^c_{i,j}(\Ft)+r^s_{i,j}(\Ft)+g^c_{i,j}+g^s_{i,j}\right]p_{i,j}\,,
\end{eqnarray}
where $r^{s/c}(F)$ are the force-dependent unbinding rates for slip ($s$) and catch ($c$) bonds, and $g^{s/c}$ are the rebinding rates setting the typical time for the formation of a new catch or slip attachment to an extracellular ligand. $\Ft$ is the total force applied to all bonds. As such, the first line of the RHS of Eqs.\,(\ref{lsme}) describes the change in $p_{i,j}(t)$ due to the unbinding of either a catch or a slip bond from a state with one additional bound bond compared to $\{i,j\}$; the second line represents rebinding of either a catch or a slip bond from a state with one fewer bound bond compared to $\{i,j\}$, and the third line represents unbinding {\em and} rebinding of either type of bond from the state $\{i,j\}$ itself. Eqs.\,(\ref{lsme}) describe a stochastic process underlying the temporal evolution of the probability distribution $p_{i,j}(t)$. Derived from it are the quantities that we will initially be most interested in, the expectation values for the total number of bound bonds $N$, and those for the numbers of bound catch ($N_{\rm c}$) and slip bonds ($N_{\rm s}$) individually
\begin{eqnarray}
N_{\rm c}(t)&\equiv& \langle i \rangle(t)=\sum_{\{i,j\}} i\, p_{i,j}(t)\nonumber \\
N_{\rm s}(t)&\equiv& \langle j \rangle(t)=\sum_{\{i,j\}} j\, p_{i,j}(t)\nonumber \\
N(t)&\equiv& \langle i+j \rangle(t)=\sum_{\{i,j\}} (i+j)\, p_{i,j}(t)\, .
\end{eqnarray}
We assume the rate of rebinding to be independent of both the applied force (because new bonds form, by definition, at zero tension) and of the type of bond. This helps simplify the initial conditioning of the system, and although it may be necessary to revisit this assumption to permit quantitative analysis we are, for the purpose of this paper, interested first in establishing the qualitative effects of mixing slip and catch bonds in adhesive clusters. Force-independent rebinding is enforced by setting
\begin{eqnarray}
g^{c}_{i,j}&=&g^{c}_i= k_0\gamma(N_{\rm ct}-i)\nonumber\\
g^{s}_{i,j}&=&g^{s}_j= k_0\gamma(N_{\rm st}-j)
\end{eqnarray}
i.e., rebinding is proportional to the instantaneous number of available, unbound bonds of the same type. Again, we simplify the system by assuming that $\gamma$ is independent of the force, and is the same for both types of bond. Of course, there is no reason for this to hold in real life; the kinetics of integrin-ligand bond formation will differ by type. 

The force-dependent unbinding rates $r^{s}(F)$ and $r^{c}(F)$ are where the differential characteristics of catch- and slip-bond manifest themselves. From now on we describe the process in terms of the total dimensionless force $\Phi=F_t/f^\star$, and define
\begin{eqnarray}
r^{c}_{i,j}(\Phi)&=&r^c_i(\Phi)\equiv i \, k_0 \, \kucb(\bar \phi)\nonumber \\
r^{s}_{i,j}(\Phi)&=&r^s_j(\Phi)\equiv j \, k_0 \, \kusb(\bar \phi)\, ,
\end{eqnarray}
where the normalized rates $\kucb$ and $\kusb$ are evaluated at the average loading force, which we obtain by assuming a uniform distribution of the total load accross all bound bonds, {\em i.e.}
\be
\bar \phi = \frac{\Phi}{i+j}
\ee
Nonuniformly distributed load may well be present in focal adhesions, and may be implemented by a spatially varying distribution of $\Phi$; again we start from the simplest scenario here. With these conventions, we derive dirctly from Eq.\,(\ref{lsme}) an evolution equation for $N(t)$, the equilibrium number of bound bonds
\be\label{eq:slipcatchev}
\DDx{}{t}N=\sum_{\{i,j\}}(i+j) \left(\DDx{p_{i,j}}{t}\right)=-\langle r^c_{i,j} \rangle+\langle g^c_{i,j} \rangle -\langle r^s_{i,j} \rangle+\langle g^s_{i,j} \rangle \,,
\ee
where the summation is over all of the possible numbers $\{i,j\}$ of bound catch- and slip- bonds in a cluster, and $\langle \rangle$ denotes averages in the distribution $p_{i,j}(t)$. Eq.\,(\ref{eq:slipcatchev}) can be split into two separate equations, describing the equilibrium number of catch $N_{\rm c}=\langle i \rangle$ and slip $N_{\rm s}=\langle j \rangle$ bonds separately. Assuming that all rate functions vary slowly around their equilibrium values, we make the mean field approximation by replacing $\langle r^{c}_{i,j} \rangle$, $\langle r^{s}_{i,j} \rangle$, $\langle g^c_{i,j} \rangle$, and $\langle g^s_{i,j} \rangle$ by the first terms in their Taylor expansions around $\{\langle i \rangle,\langle j \rangle\}$: $\langle r^{c}_{i,j} \rangle \approx  r^{c}_{\langle i \rangle,\langle j \rangle}$, $\langle g^{c}_{i,j} \rangle \approx  g^{c}_{\langle i \rangle,\langle j \rangle}$ etc. This transforms Eqs.\,(\ref{eq:slipcatchev}) into the following coupled system
\begin{equation} \label{eq:system}
\left\{
\begin{alignedat}{4}
\DDx{}{t}{N_{\rm c}} = & -N_{\rm c} \kucb \left(\frac{\Phi}{N_{\rm c}+N_{\rm s}}\right)+\gamma (N_{\rm ct} - N_{\rm c})\, \\
\DDx{}{t}{N_{\rm s}} = & -N_{\rm s} \kusb \left(\frac{\Phi}{N_{\rm c}+N_{\rm s}}\right)+\,\gamma (N_{\rm st} - N_{\rm s})\, .
 \end{alignedat}
 \right. 
\end{equation}
Here the time $t$ is actually the nondimensionalized time $t k_0$, but we may set $k_0=1$s without losing generality. Note, also, the nature of the coupling: In our model, the different types of bonds are aware of each other only through the shared total force $\Phi$. At equilibrium, the RHS of both equations in the system above vanish. At zero overall force, the equations fully decouple. The number of bound slip bonds becomes independent of $\rho_{\xi}$. For general forces, the coupled system \ref{eq:system} has two solutions for each value of force. One of the solutions is unstable, the other solution corresponds to the local equilibrium and is stable. These two solution branches are readily obtained by direct numerical solution of Eqs.\,(\ref{eq:system}), with the RHS's equated to zero. 

As shown in Fig. \ref{Fig2}, at the equilibrium level the effect of mixing catch and slip bonds is that slip bonds provide most of the adhesion at low forces, while the catch bonds take over at intermediate and high forces. This is a marked increase in functionality over having just catch bonds; while these are able to stabilize adhesions at high forces they must pass through an extended, weakly bound regime to get there. Mixed catch-slip adhesion clusters always have an appreciable number of the integrins bound and as such provide stability at all force levels. We now compare these numerical solutions to the results of stochastic simulations of the mixed bonds system.
\section{Stochastic simulations of mixed clusters: Equilibrium bond numbers} 

\begin{figure}[th]
\includegraphics[width=\columnwidth]{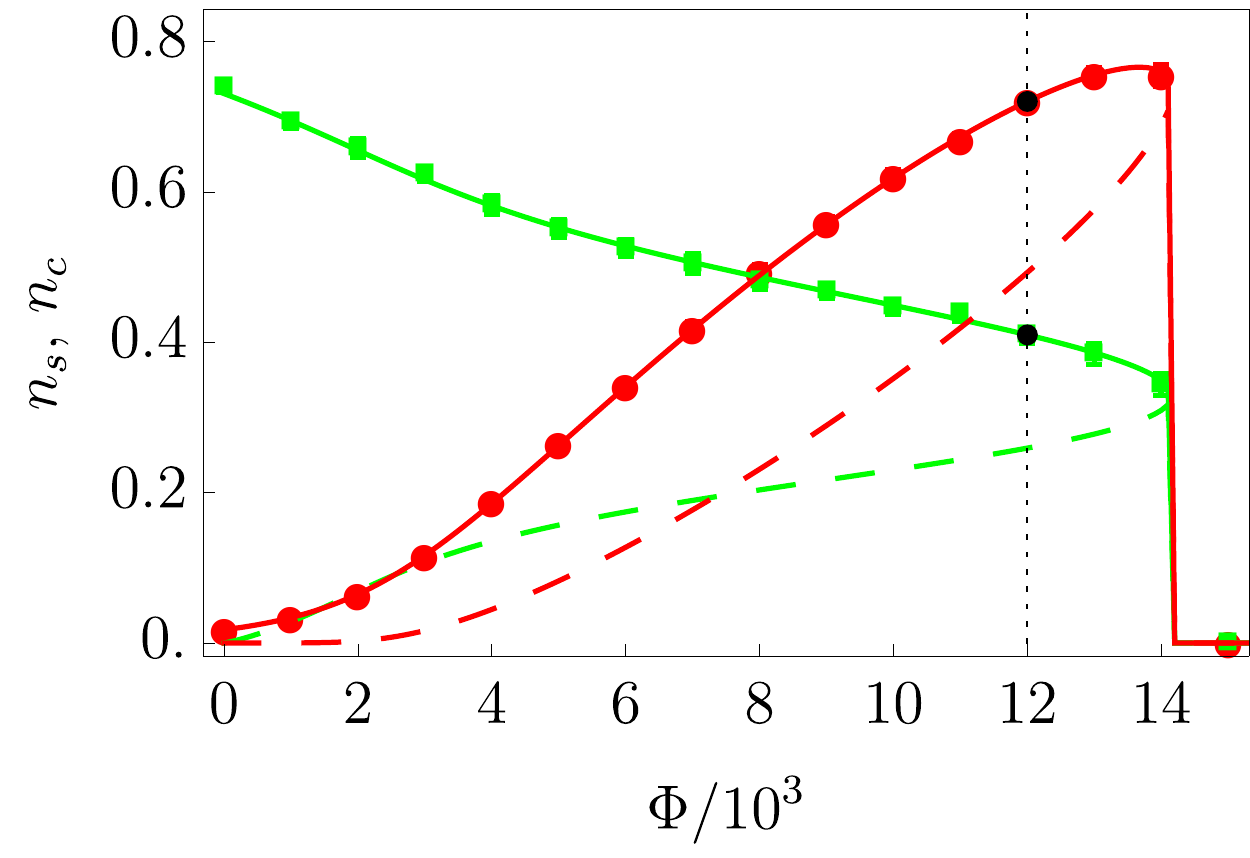}\caption{\label{Fig2}Relative fraction of closed catch and slip bonds as a function of the force in a cluster of 2048 catch and 2048 slip bonds. Orange/blue points correspond to simulation results for catch/slip bonds, starting from all bonds closed for zero force. Pink/green lines - deterministic solution, obtained by solving equations \ref{eq:system}, for experimentally derived catch bond force-lifetime curves and slip bonds with $\rho_{\xi}=3.8$ and $u_{\rm sb}=1$. Rebinding rate for both catch and slip bonds is $\gamma=1$. The vertical line at $\Phi=12\times 10^3$ is the total force at which we simulate the dynamics for Fig. \ref{Fig3}; the two black dots where this line intersects the stable branches for catch and slip bonds represent the predicted equilibrium binding fractions.}
\end{figure}

\begin{figure}
\includegraphics[width=\columnwidth]{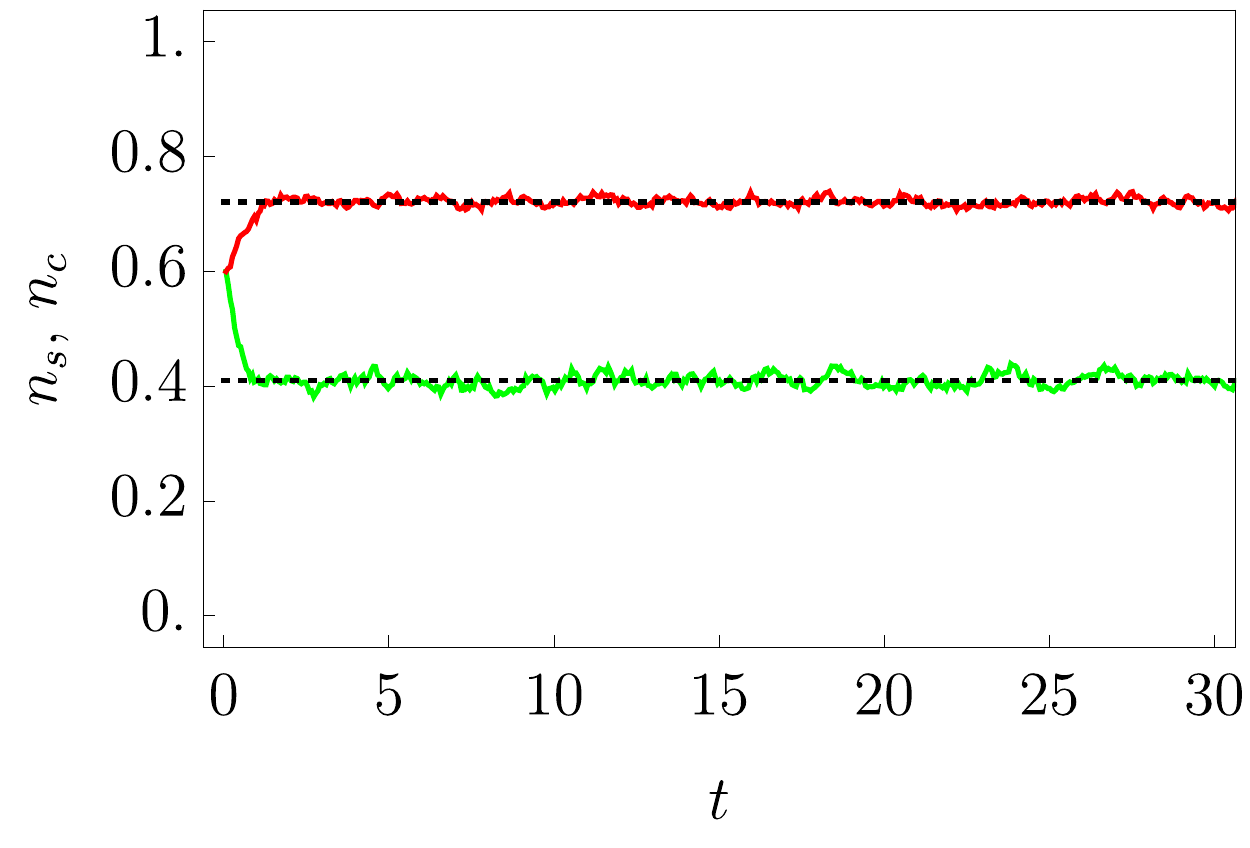}\caption{\label{Fig3}Relative fraction of closed catch and slip bonds as a function of time for $\Phi=12\times 10^3, \rho_{\xi}=3.8$ and $u_{\rm sb}=1$, $\gamma=1$, 2048 catch bonds and 2048 slip bonds. Bound fractions for both were initialized at $0.6$. The red and green lines represent the evolution of the fraction of bound catch (red) and slip (green) bonds over time. The horizontal black dotted lines represent the predicted equilibrium values for these fractions according to Eq.\,(\ref{eq:system}). These values also correspond to the black dots in Fig. \ref{Fig2}.}
\end{figure}

While the mean field approximation can teach us something about equilibrium behavior and expectation values, it says nothing about the dynamic behavior and in particular is not able to address the lifetime of the stable state. As we have demonstrated in earlier work, the mechanism for cluster unbinding is fluctuation-driven, what we have called the stable solution branch is actually a metastable branch and a sufficiently large bond number fluctuation---which will come along at some point---prompts unbinding of the entire cluster. In order to address the lifetime of mixed clusters, we therefore turn to stochastic simulations, for which we use the Gillespie algorithm \cite{gillespie}. We initiate the system at a certain total number of bound bonds of each type, and specify the cluster composition (total numbers of available catch and slip bonds). The choice of the initial value of bound bonds determines the typical evolution of the simulation, in the sense that in order to reach the (meta)stable solution branch the initial values must be chosen within the basin of attraction of that branch. A typical simulation allows us to compute the typical evolution of the number of bound catch and slip bonds with time, as Fig. \ref{Fig2} demonstrates. The solid lines are the equilibrium predictions from Eqs.\,(\ref{eq:system}), and indeed the system is seen to converge onto the predicted values after a brief equilibration period. For this particular choice of parameters, the cluster is stable over the entire time of the simulation. However, the stochastic simulations also capture cluster unbinding, as is shown in Fig. \ref{Fig3}, where an initially stabilized cluster unbinds after a spontaneous supercritical bond number fluctuation.  Repeating these simulations multiple times, for different total forces and different parameter values, we collecting statistics on both the average values of the number of bound bonds of each type, and the lifetime of the composite cluster. Fig. \ref{Fig3} shows that, as predicted by the mean-field model, the average relative numbers of bound catch ($n_s=\langle N_{\rm s}(t) \rangle_t/N_{\rm st}$) and slip ($n_c=\langle N_{\rm c}(t) \rangle_t/N_{\rm ct}$) bonds in a stable adhesive cluster follows the expected behavior, and that catch and slip bonds preserve their tendencies even when coupled to each other via the force applied to a composite cluster. The number of catch bonds still peaks at some finite forces, while the equilibrium fraction of bound slip bonds decreases monotonically with increasing force. In measuring these average bound bond numbers, we take into account only the times during which a stable adhesion is present; should the cluster unbind we stop measuring. Thus, what this simulation is bearing out is that the composition of stably adherent clusters is reliably predicted by Eqs.\,(\ref{eq:system}).

\section{Stochastic simulations of mixed clusters: Cluster lifetimes} 

\begin{figure}
\includegraphics[width=\columnwidth]{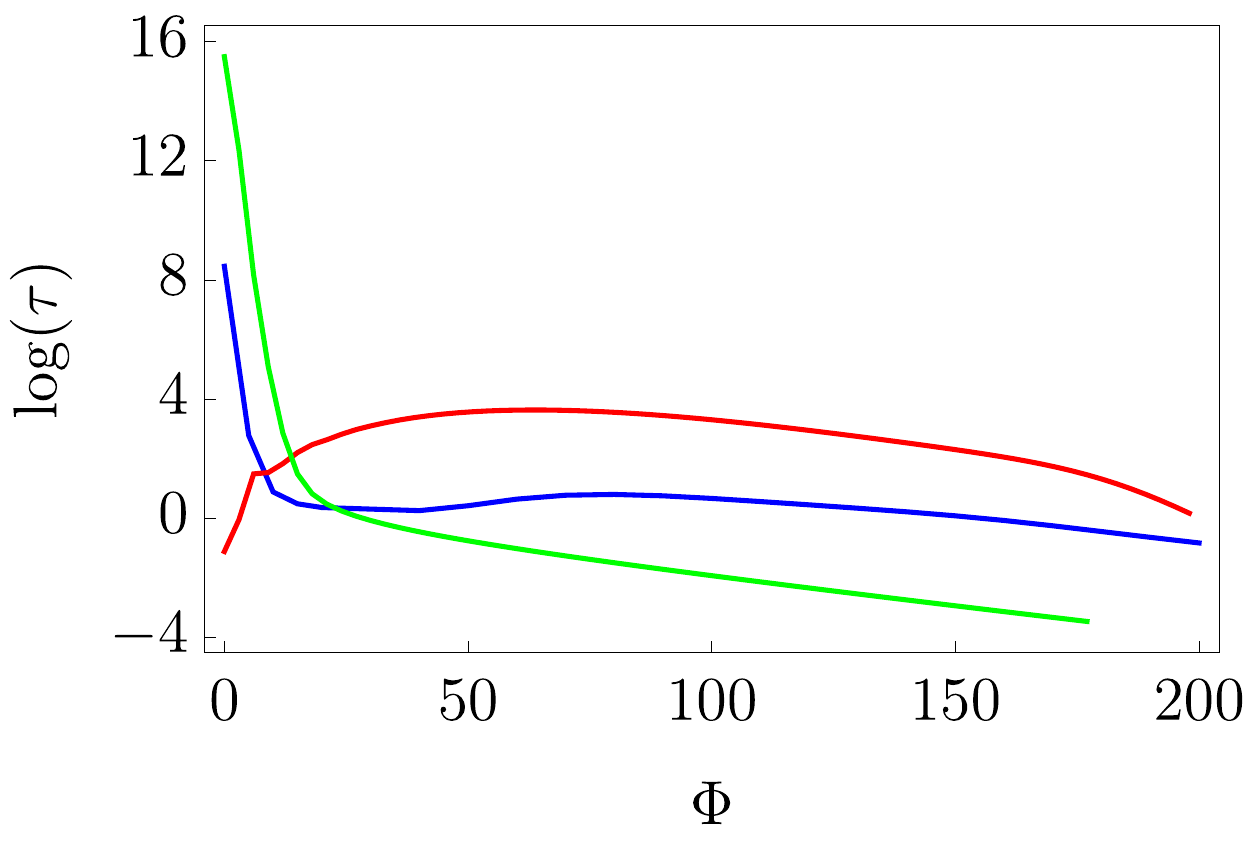}\caption{\label{Fig4}Comparison between the lifetimes of a cluster consisting of 100 catch bonds (red curve, parameters: $\rho_{\xi}=1$, $\gamma=0.2$,$u_{sb}=1$), 100 slip bonds (green curve, parameters: $\rho_{\xi}=1$,$\gamma=0.2$,$u_{sb}=1$) and a cluster containing 50 slip bonds and 50 catch bonds (blue curve, same parameters as the pure systems). }
\end{figure}

With the force-dependent numbers of bound bonds and their partitioning between catch and slip now clear, we may ask what functional advantage, if any, the presence of both types of bonds offers over only a single species of integrin. Is it true, that the increased presence of bound bonds (mostly slip) at low forces translates into increased lifetimes in this region, and is this providing additional and previously missing low-force stability? Our stochastic simulations allow us to measure the lifetime of a mixed cluster, and compare it to the lifetimes of clusters containing only catch, or only slip bonds. Representative results are collected in Fig. \ref{Fig4}. 

\begin{figure}
\includegraphics[width=\columnwidth]{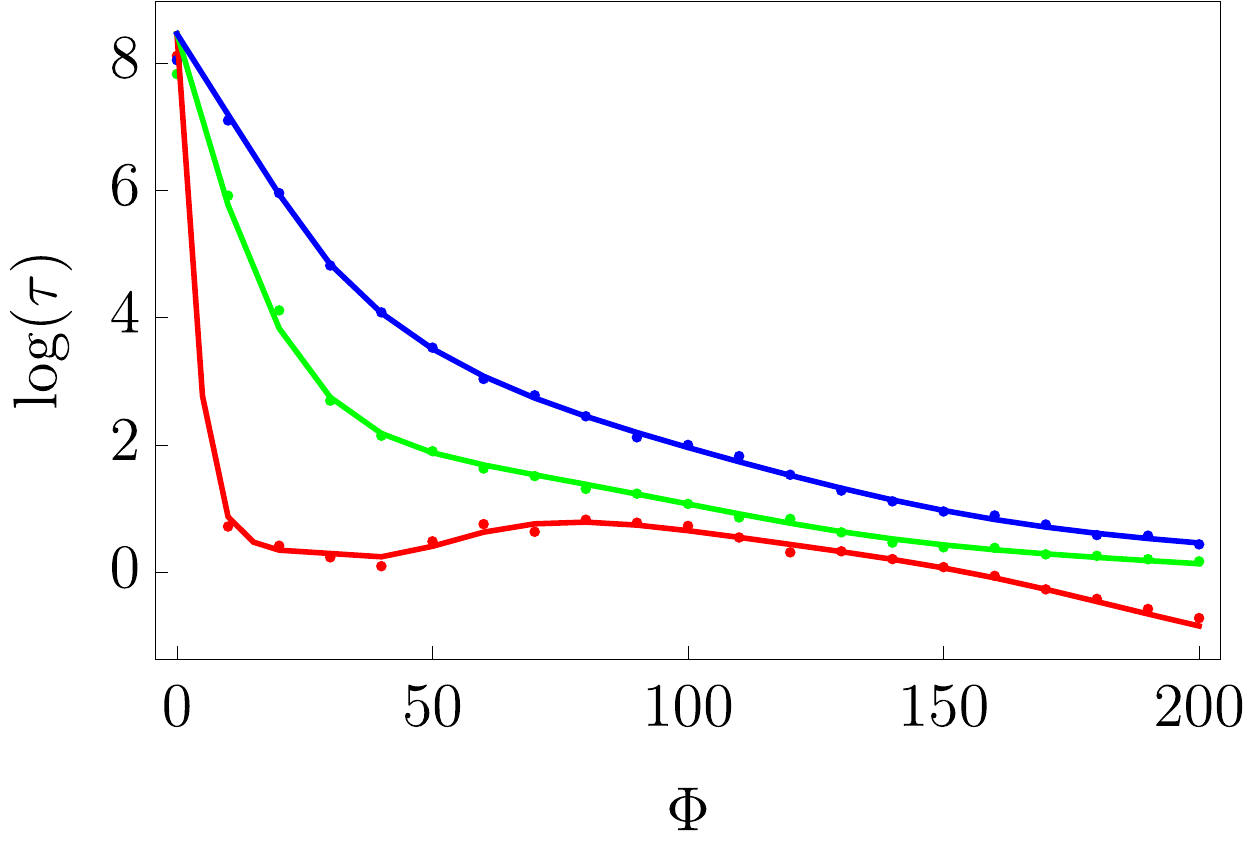}\caption{\label{Fig5}Lifetime of a cluster consisting of 50 catch and 50 slip bonds depending on force. Each of the points represent the result of 100 simulation trajectories started  out from $N_{\rm c}=25$ and $N_{\rm s}=25$ after $4\cdot 10^6$ steps each, and $\rho_{\xi}=1$ (red) $\rho_{\xi}=3.8$ (green), $\rho_{\xi}=6.6$ (blue), solid line of matching color correspond to $T_{25,25}$ - the solution of Eq.\,(\ref{lifetimes}).}
\end{figure} 

 As Fig. \ref{Fig4} illustrates, mixing catch- and slip bonds provides additional functionality compared to either of the two single-component systems. At low forces, the slip bonds provide initial stability to a nascent cluster compared to a catch-only cluster. This eliminates the weakly bound low-force regime from the pure catch system; the slip bonds ensure immediate and effective adhesion. As the force rises, and slip bonds are  gradually replaced by catch bond integrins the behavior of the entire cluster increasingly reflect their high-force stabilizing effect; the blue curve is above the green curve. The advantage of mixing is thus obvious from the mean lifetime; the catch bonds in the mixed cluster provide additional stability and a far higher threshold force for unbinding at high forces compared to pure slip bond systems, while the slip bonds provide greatly enhanced stability at lower forces.
 
 That the lifetime of the mixed cluster is nowhere longer than either the pure catch or the pure slip system shouldn't come as a surprise; in the regimes where the behavior of one type of bonds dominates this behavior is always going to be diluted to some extent by the presence of the other, subdominant bond type. We speculate that the overall improvement of both low- and high force stability takes precedence over further increases in lifetime at one particular force regime. 
 
Fig. \ref{Fig4} also suggests a particular sequence to the dynamics of integrin recruitment to developing focal adhesions. As the tension builds in the stress fiber attached to the focal adhesion, the system travels along the $\Phi$-axis. Based on our mixed-cluster model, we suggest this phase of tension-buildup drives a shift in FA composition, or at least in the partitioning of those integrins that are bound to the substrate. Younger focal adhesions benefit most from bound slip-type bonds, whereas mature focal adhesions will rely more on catch bonds.

The partitioning of bound bonds will be exceedingly difficult to measure directly. Their complement---the unbound bonds---may well be a better target to validate the predicted behavior. In the following sections, we detail how careful observation of the diffusive behavior of both bond types inside the FA may reveal a force-dependent compositional shift.
 
\section{Lateral diffusion of catch and slip bonds in the adhesive zone}
Why should the force-dependent composition of a mixed-cluster adhesion affect the mean diffusivity of integrins inside a FA? To see this, consider an area densely covered with integrins of both types, some bound and some unbound. In-plane hopping of one integrin to a neighboring site then requires it to exchange places with a neighbor that is also not bound, and therefore also free to move. An abundance of bound bonds, which are immobilized by their connection to the ECM, in this environment reduces the opportunities for such hops, and thus strongly suppresses the diffusivity of unbound integrins.  Indeed, single-protein tracking experiments \cite{Giannone} report clear changes in the diffusivity depending on the applied tension. To model the diffusion, we include now the spatial distribution of integrins in our model by putting the integrins on a square lattice, with lattice spacing $\lambda$.

\begin{figure}[thb]
\includegraphics[width=\columnwidth]{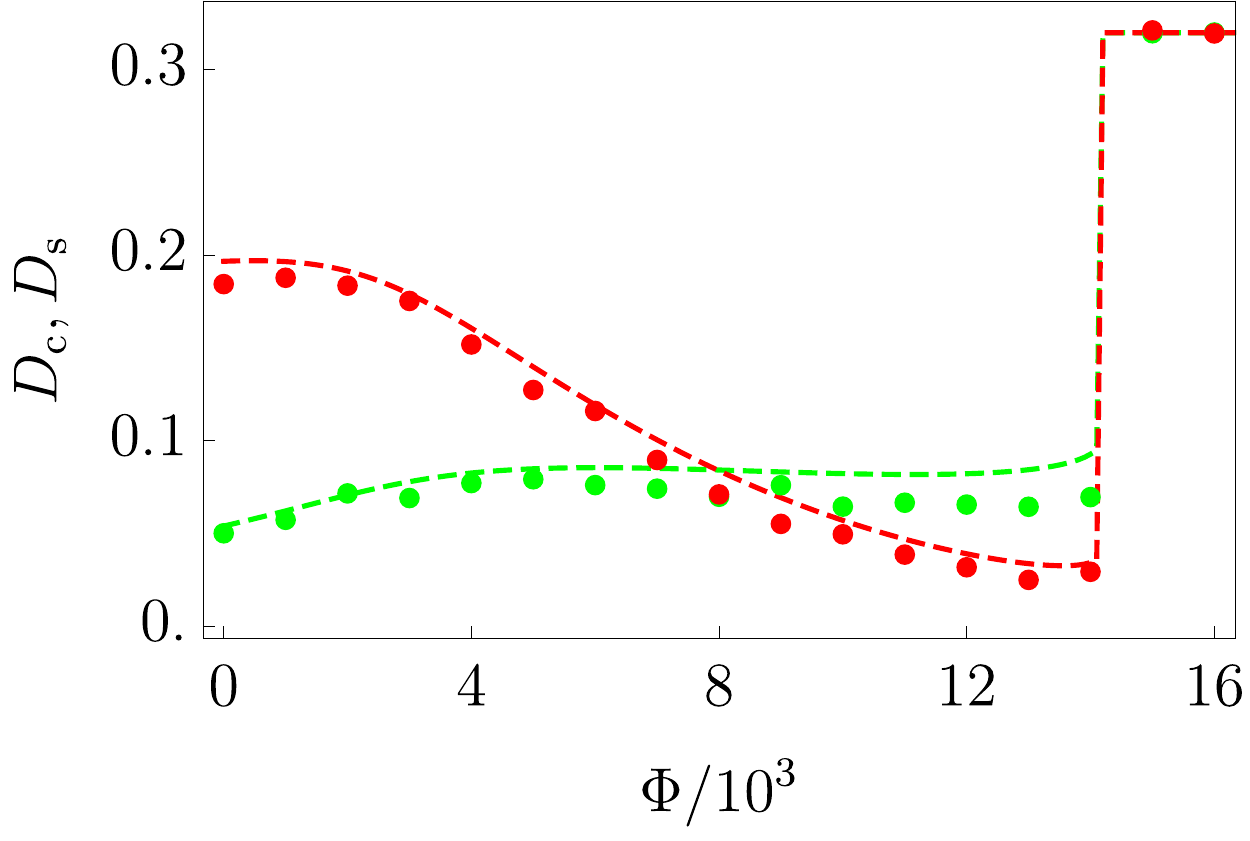} \caption{\label{Fig6}  Mean diffusivity of catch and slip bonds as a function of the force exerted on the cluster. Parameter values: $\rho_{\xi}=3.8$ and $u_{\rm sb}=1$, $\gamma=1$. Dots are simulation results, solid lines are calculated from Eq.\,(\ref{Diff}).} 
\end{figure}

In these simulations, the binding and unbinding behavior is as it was before in the Gillespie approach, but now we add as a potential update move the exchange of position between two neighbouring lattice sites provided both are unbound. In such a simulation, the diffusion coefficient $D$ may be computed following \cite{haus1987diffusion} as the coefficient of proportionality between the mean residence time at a lattice site $\langle t_{\rm res}\rangle$ and the squared lattice spacing:
\be \label{DiffusionKerr}D= \frac{\lambda^2}{2 d \langle t_{\rm res}\rangle}\,,
\ee
where $d=2$ is the dimensionality of the lattice. For a single, unbound integrin on an otherwise empty lattice, the transition rate $r_0$ for hopping between neighboring sites according is set to 
\be
r_0=\frac{D_0}{\lambda^2}\,;
\ee
we shall refer to $D_0$ as the free diffusion constant. In a simulation run with many binding and unbinding integrins of both types then proceeds as follows: neighboring unbound bonds to exchange sites with a rate $r_0$. To be able to put some actual numbers on the quantities we compute, we choose the lattice spacing $\lambda$ such that the total density $\rho_{\rm tot}$ of integrins matches the value reported in \cite{RocaCusachs2014}, setting $\lambda=\sqrt{\rho_{\rm tot}}\approx 20$ nm. The free diffusion constant is set to $D_0=0.32$ Subject to the rule that exchanges are only permitted if both neighbors are unbound, we measure how long each bond spends at a single lattice site before moving to the other site. Averaging over all bonds of a single type (catch or slip) we compute the mean residence time, $\langle t_{\rm res,c/s}\rangle$, from which according to Eq.\,(\ref{DiffusionKerr}) for a 2D system the diffusion coefficients may be computed as:
\be
D_{c/s}=\frac{\lambda^2}{4\langle t_{\rm res,c/s}\rangle}\,.
\ee
The diffusion coefficient for either bond type, in a system with a given number of catch- and slip bonds is determined by two factors: how many bonds of a given type are able to move ({\em i.e.}, are unbound), and how many unbound neighbors of either type are in the direct vicinity. Fig. \ref{Fig6} shows the resulting behavior. The dots in this figure represent simulation data and show that the changing composition of the cluster, as the force rises, is indeed reflected directly in the diffusive behavior of the free integrins. Initially, the mobility of the catch bonds is considerably higher, reflecting the fact that many of them are not yet bound and thus able to diffuse. The slip bonds, in contrast, are mostly bound and thus a large fraction of them is immobile. As the force increases, this picture is reversed and while the slip bonds are, on average, becoming increasingly mobile more and more catch bonds are becoming bound and immobile.

 This simple physical picture can be summarized in the following formula for the effective, force-dependent diffusion coefficient of catch and slip integrins in adhesion sites densely covered in integrins
\be \label{Diff}
D_{\rm c/s}(\Phi)=D_0 \biggl(1\!-\!n_{c/s}(\Phi)\biggr)\biggl[1\!-\!\frac{N_{\rm ct}\!n_{c}(\Phi)}{N_{\rm ct}+N_{\rm st}}-\frac{N_{\rm st}n_{s}(\Phi)}{N_{\rm ct}+N_{\rm st}}]\biggr]\,,
\ee
where $n_{c}(\Phi)$ and $n_{s}(\Phi)$ are the fraction of bound catch- or slip- bonds, respectively. The term between round brackets accounts for the availability of nonbound bonds, the term between square brackets accounts for the availability of nonbound neighbours. The predictions of Eq.\,(\ref{Diff}), after plugging in the equilibrium values of $n_{c}(\Phi)$ and $n_{s}(\Phi)$ computed earlier, are graphed with solid lines in Fig. \ref{Fig6}, confirming the agreement with our simulations. Comparing Fig. \ref{Fig6} with Fig. \ref{Fig2} confirms the intuitive correspondence between diffusivity and the bound/unbound fractions of both species.

\section{Discussion}
We have studied the behavior of adhesive clusters composed of a mixture of catch and slip bonds. Our results show, that such mixed clusters provide increased functionality over either of the two pure systems---the bonds, in fact, complement each other in the sense that the addition of slip bonds provides additional stability at low forces to purely catch systems, and the addition of catch bonds provides increased load-bearing capacity and strength at higher forces. While our model does not include direct interactions between the two types of bonds, they {\em do} interact indirectly, via the shared force. 

As a result of this indirect, nonlinear coupling between the bond types, the fractions of bound bonds, for both species, change as the force is increased. Our model therefore suggests, that the two types of bonds not only play different roles within a composite cluster, but that they are also differentially engaged depending on the applied force. Because the force exerted at a given focal adhesion increases as the adhesion matures, this implies that the engagement (or activation) of different integrin species automatically becomes organized in time, with early adhesions featuring mostly bound slip bonds and late-stage adhesion featuring more adherent catch bonds.

In experiments, this differential engagement will be exceedingly difficult to quantify or even image directly, because all of this may happen even against a background of constant overall focal adhesion composition. What changes over time are the fractions of bonds of either type that are actually {\em  bound} to the ECM. To circumvent this difficulty, we suggest to measure, rather, the average diffusivity of the different types of integrins inside a focal adhesion which we find to report directly on their instantaneous activation (engagement). Moreover, {\em changes} in these diffusivities might be used to assess force-dependent changes in the contributions of these different species. While, to be sure, this is still by no means straighforward it has actually been demonstrated in previous experiments \cite{Giannone}. Our results show, that similar measurements executed at different times can compare nascent, early and mature focal adhesions and have the potential to verify the differential engagement of various integrin types during adhesion. Again, we stress that engagement and concentration are two distinct quantities; the presence of an integrin does not imply its state of activation.   

While it is most certainly oversimplifying the spectacular biophysics of the focal adhesion our model is a first attempt to quantitatively assess the benefits of complexity and redundancy in cellular adhesion. We find, that even with two only species such benefits are readily identified, may be intuitively understood and modeled, and that the evolution of the system is robustly self-organized--- encoded through physical, statistical-mechanical principles rather than specific biochemical regulation. Experiments well within reach of the current state-of-the-art should be able to confirm some of the predictions we make here.
\\
\begin{acknowledgements}
This work was supported by funds from the Netherlands Organization for Scientific Research (NWO-FOM) within the program on Mechanosensing and Mechanotransduction by Cells (FOM-E1009M). We thank Prof. Ulrich Schwarz, Prof. Erik Danen, Dr. Thorsten Erdmann and Dr. Emrah Balcioglu for valuable discussions.
\end{acknowledgements}

\bibliography{references}

\appendix
\section{Analytical calculation of the lifetime of a mixed cluster}

\begin{figure}[th]
\includegraphics[width=\columnwidth]{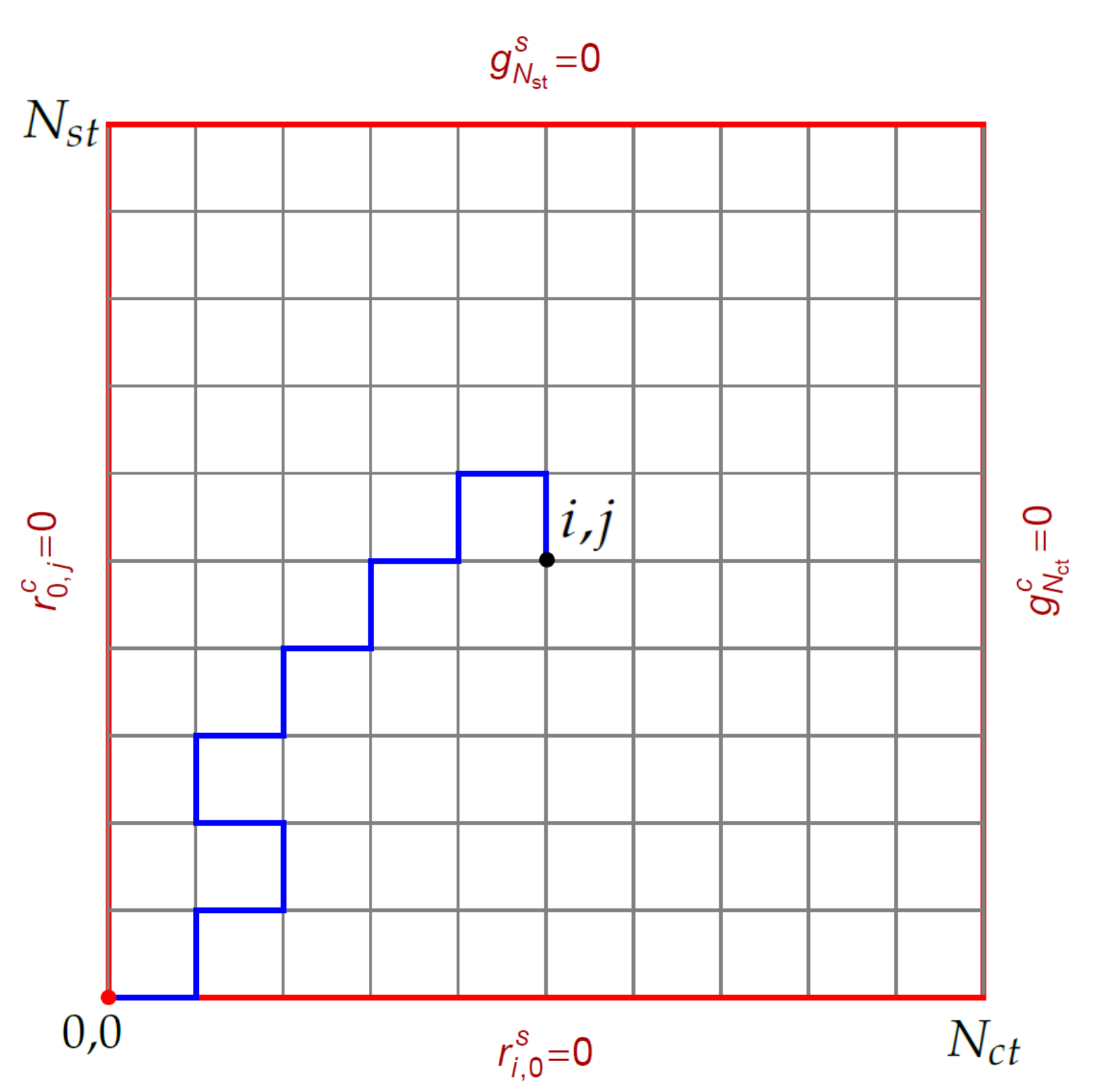}\caption{\label{Fig7}Sketch of the configurational space that cluster with two types of bonds explores. The parameters of the system are the number of bound catch bonds (along the $x$-axis) and the number of bound slip-bonds (along the $y$-axis). All unbinding pathways correspond to trajectories that end up in the origin at the lower left corner. The example trajectory of unbinding (blue line) starts from $(i,j)$ bound bonds(black point) and ends at an absorbing  boundary at $(0,0)$ (the red point); it is subject to reflecting boundaries along the red lines. The trajectory is confined to be inside the phase space at all times when $0 \leqslant i \leqslant N_{\rm st}$ and $0 \leqslant j \leqslant N_{\rm ct}$.}
\end{figure}

The time $T_{i,j}$, that it takes a cluster of $i$ bound catch and $j$ bound slip bonds to reach the point where all catch and slip bonds are unbound, obeys a recursive equation which may be derived using the methods set out in \cite{VanKampen}. This relation reads
\begin{eqnarray} \label{lifetimes}
&T_{i, j}&=T_{i+1,j}\frac{g_i}{g_i\!+\!g_j\!+\!r^c_{i,j}\!+\!r^s_{i,j}}+\nonumber\\ &+&T_{i,j+1}\frac{g_j}{g_i\!+\!g_j\!+\!r^c_{i,j}\!+\!r^s_{i,j}}\!+\!T_{i,j-1}\frac{r^s_{i,j}}{g_i\!+\!g_j\!+\!r^c_{i,j}\!+\!r^s_{i,j}}+\nonumber\\ &+&T_{i-1,j}\frac{r^c_{i,j}}{g_i\!+\!g_j\!+\!r^c_{i,j}\!+\!r^s_{i,j}} \!+\!\frac{1}{g_i\!+\!g_j\!+\!r^c_{i,j}\!+\!r^s_{i,j}}\,,
\end{eqnarray}
with $g$ and $r$ the binding and unbinding rates as defined in the main text. The last term in Eqs.\,(\ref{lifetimes}) corresponds to the time that it takes to leave state ${i,j}$ to any of its neighboring states in configurational space, and the first four terms represent the lifetimes of those four neighboring states, multiplied by the transition probabilities to those states. Writing this out for all possible combinations of catch- and slipbonds, one obtains $N_{\rm c} \times N_{\rm s}$ equations for $T_{i,j}$. This system of coupled algebraic equations is to be solved subject to a number of boundary conditions: 
\begin{eqnarray}
T_{0,0}&=&0 \quad:\quad \text{absorbing boundary},\label{Boundaries1} \\
T_{-1,0}&=&0 \quad:\quad\text{no negative }i, \label{Boundaries2}\\
T_{0,-1}&=&0 \quad:\quad \text{no negative }j,\label{Boundaries3} \\
g^s_{N_{\rm st}}&=&0 \quad:\quad \text{reflecting boundary},\label{Boundaries4} \\
g^c_{N_{\rm ct}}&=&0 \quad:\quad \text{reflecting boundary}, \label{Boundaries5}\\
r^s_{i,0}&=&0 \quad:\quad \text{reflecting boundary}, \label{Boundaries6}\\
r^c_{0,j}&=&0 \quad:\quad \text{reflecting boundary}.\label{Boundaries7}
\end{eqnarray}
Eq.\,(\ref{Boundaries1}) reflects that the cluster does not rebind after all its bonds are unbound. Eqs.\,(\ref{Boundaries2}) and (\ref{Boundaries3}) express the condition that the number of bound bonds is never negative. Eqs.\,(\ref{Boundaries4}) and (\ref{Boundaries5}) take care that the cluster can not rebind more bonds than are available, and finally Eqs.\,(\ref{Boundaries6}) and (\ref{Boundaries7}) take care that the rupture rates vanish when no bonds of each type are bound. .

The analytical expression for the solution of system \ref{lifetimes} is quite bulky, and cannot be expressed in a compact from for each of the $T_{i,j}$. However, Eq.\,(\ref{lifetimes}) is straightforwardly solved for a given total number of catch and slip bonds. These solution are graphed in Fig. \ref{Fig5}, where we calculate the lifetime of a cluster consisting of 50 catch bonds and 50 slip bonds with various parameters and confirm the analytical outcome by comparing to stochastic simulations.

\end{document}